\documentclass[letterpaper]{article} 
\usepackage{aaai25}  
\usepackage{times}  
\usepackage{helvet}  
\usepackage{courier}  
\usepackage[hyphens]{url}  
\usepackage{graphicx} 
\usepackage{natbib}  
\usepackage{caption} 
\usepackage{algorithm}
\usepackage{algorithmic}

\usepackage{newfloat}
\usepackage{listings}
\DeclareCaptionStyle{ruled}{labelfont=normalfont,labelsep=colon,strut=off} 
\floatstyle{ruled}
\newfloat{listing}{tb}{lst}{}
\floatname{listing}{Listing}
\usepackage{enumitem}
\usepackage{booktabs}
\usepackage{xcolor}
\usepackage{amsmath}
\usepackage{tabularx}
\usepackage{subcaption}

\newcommand{\mff}{$\mathtt{M_{50}}$}
\newcommand{\mtw}{$\mathtt{M_{25}}$}
\newcommand{\mtn}{$\mathtt{M_{10}}$}

\newcommand{\revised}[1]{\textcolor{black}{#1}}

\newcommand{\treatmentcomments}{TC}
\newcommand{\treatmentundesired}{TU}
\newcommand{\treatmentrank}{TR}
\newcommand{\treatmentcommentsweak}{TC-W}
\newcommand{\treatmentcommentsstrong}{TC-S}
\newcommand{\treatmentundesiredweak}{TU-W}
\newcommand{\treatmentundesiredstrong}{TU-S}
\newcommand{\treatmentrankweak}{TR-W}
\newcommand{\treatmentrankstrong}{TR-S}

\usepackage{newfloat}
\usepackage{listings}
\floatstyle{ruled}
\newfloat{listing}{tb}{lst}{}
\floatname{listing}{Listing}

\title{Examining Algorithmic Curation on Social Media:\\ An Empirical Audit of Reddit's r/popular Feed}

\author{
    Jackie Chan,
    Fred Choi,
    Koustuv Saha,
    Eshwar Chandrasekharan
}

\affiliations{
    University of Illinois Urbana-Champaign\\
    \{jackiec3, fc20, ksaha2, eshwar\}@illinois.edu
}

\usepackage{bibentry}

\begin{document}

\maketitle

\begin{abstract}
    Platforms are increasingly relying on algorithms to curate the content within users' social media feeds. However, the growing prominence of proprietary, \textit{algorithmically curated feeds} has concealed \textit{what factors influence the presentation} of content on social media feeds and \textit{how that presentation affects user behavior}. This lack of transparency can be detrimental to users, from reducing users' agency over their content consumption to the propagation of misinformation and toxic content. To uncover details about how these feeds operate and influence user behavior, we conduct an empirical audit of Reddit's algorithmically curated trending feed called \textit{r/popular}. Using 10K r/popular posts collected by taking snapshots of the feed over 11 months, \revised{we find that recent comments help a post remain on r/popular longer and climb the feed. We also find that posts below rank 80 correspond to a sharp decline in activity compared to posts above. When examining the effects of having a higher proportion of undesired behavior---i.e., moderator-removed and toxic comments---we find no significant evidence that it helps posts stay on r/popular for longer. 
Although posts closer to the top receive more undesired comments, we find this increase to coincide with a broader increase in overall engagement---rather than indicating a disproportionate effect on undesired activity.}
\revised{The} relationship\revised{s} between algorithmic rank and engagement highlight the extent to which algorithms employed by social media platforms essentially determine which content is prioritized and which is not. We conclude by discussing how content creators, consumers, and moderators on social media platforms can benefit from empirical audits aimed at improving transparency in algorithmically curated feeds.

\end{abstract}

\section{Introduction}

Social media platforms are flooded with immense amounts of new content every day. This constant stream of content has led to the reliance on \textit{algorithms} to \textit{curate} the content in users' social media feeds.  \textit{Algorithmic curation} is defined as the process of ``organizing, selecting, and presenting subsets of a corpus of information for consumption''~\cite{rader-gray-folk-theories}. In recent times, the most prominent examples of algorithmic curation are on short-form video platforms like TikTok, YouTube Shorts, and Instagram Reels. However, less thought of are the ``hot'' and ``popular'' feeds on multiple platforms (e.g., GitHub, Reddit, X/Twitter) that also use algorithms to curate what is ``trending''~\cite{trending-is-trending}.

Despite the prominent usage of algorithmic curation on social media, uncovering \textit{what factors influence curation algorithms} is challenging. This is because curation algorithms are proprietary, preventing those on the outside from knowing how these algorithms work, thus protecting companies' intellectual property. Additionally, algorithmic curation on social media is mostly done for specific users, i.e., what these algorithms recommend changes for each individual user. The prominence of personalized recommendations on social media presents two challenges: (1) understanding how these systems affect users more broadly, and (2) collecting enough data to support studies into these systems. To alleviate these challenges, we present a study that examines the trending feed on Reddit called r/popular.

\begin{figure}
    \centering
    \includegraphics[width=\linewidth]{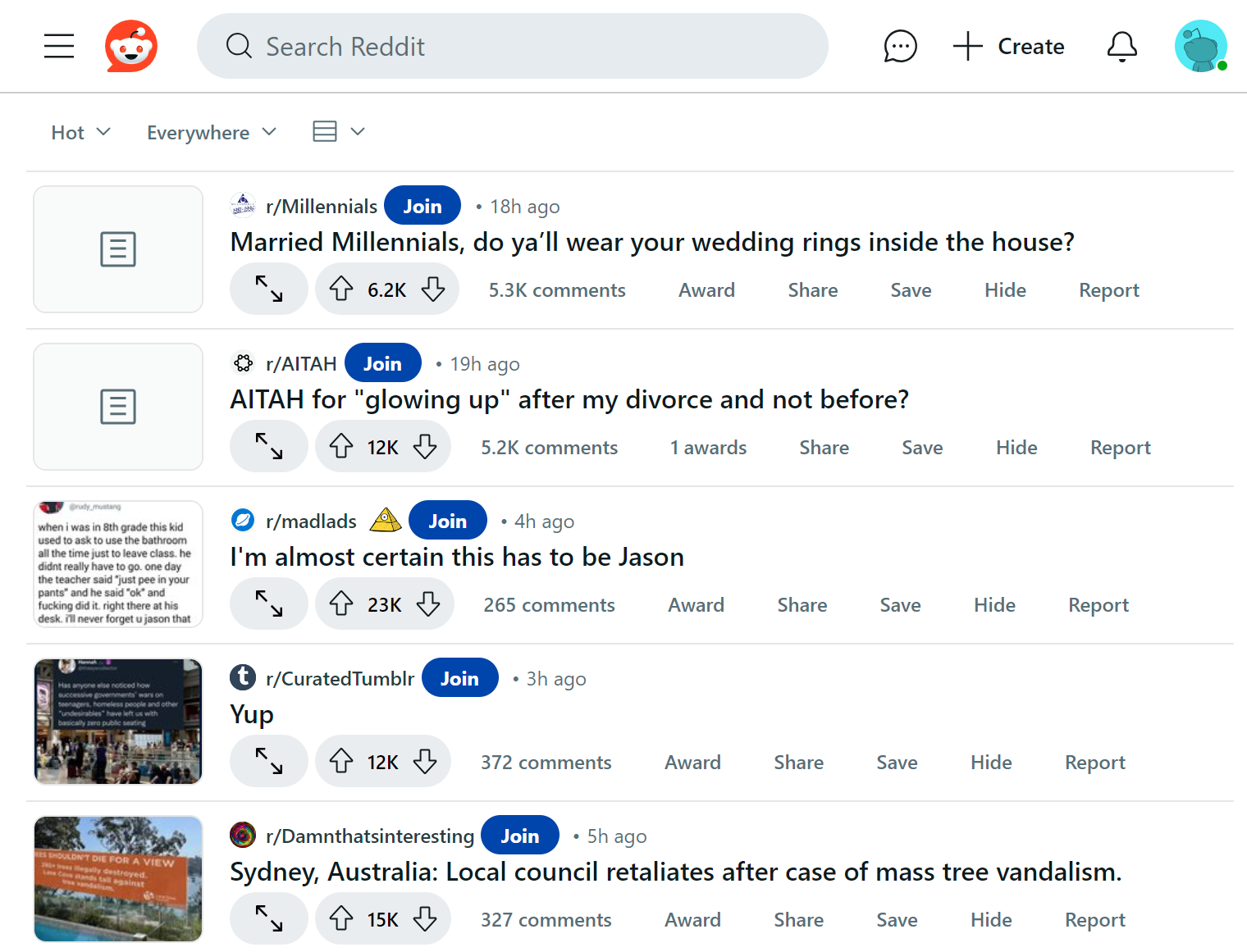}
    \caption{A snapshot of the r/popular feed on Reddit.}
    \label{fig:popular-feed}
\end{figure}

\revised{Before describing r/popular in further detail, we emphasize the importance of transparency in algorithmic curation---particularly ranking. Ranking shapes online behavior~\cite{watts-music-markets, reranking} and is often interpreted as a signal of quality~\cite{joachims-clicks}. This gives platforms significant influence over what content receives attention and how users engage with it~\cite{serps-apis}, with downstream effects on political beliefs~\cite{seme-politics-candidates, political-search-bias} and the ability to assess truth online~\cite{seme-medicine}. We extend research on algorithmic transparency by examining \textit{what drives} and \textit{is driven by} content ranking in social media feeds.}

\subsection{Reddit's r/popular Feed}

The r/popular feed\footnote{\textit{https://reddit.com/r/popular}} on Reddit is one of the default feeds that is available to all users, with and without an account. The posts on the r/popular feed come from nearly all consenting communities on the platform---however, there are some subreddits, particularly ``not safe for work'' (NSFW) communities, that are omitted. The r/popular feed is a \textit{ranked} list (see Figure~\ref{fig:popular-feed}) where the rankings are based on a calculated ``hot'' score (likely based on recency, comment rate, and upvote score) as well as some measures for post and discussion quality.\footnote{\textit{https://reddit.com/9n3ix9}} At a high level, the r/popular feed consolidates the most active posts on Reddit from numerous subreddits and serves as the front page to the platform.

Studying Reddit's r/popular feed comes with several benefits: (1) it is highly visited, (2) it is available to all users, and (3) it orders the posts consistently for all users, making it impactful for broad population. Its prominence also allows us to collect significant amounts of data to uncover what goes into Reddit's ranking decisions. Thus, analyzing the r/popular feed can provide key insights into how algorithmically curated feeds work and how algorithmic ranking decisions can impact user behaviors on platforms like Reddit.

\subsection{Our Contributions}
In this paper, we conduct an algorithmic audit of r/popular with two key objectives. The first is to understand what factors influence algorithmic ranking on r/popular. The second is to quantify how those decisions, specifically the feed's ranking decisions, affect the engagement on posts. Toward these goals, we ask three research questions.

\begin{enumerate}[label=(RQ\arabic*), leftmargin=1.25cm]
    \item What factors \revised{are associated with} how long a post stays on the r/popular feed (i.e., the post's tenure on the feed)?
    \item What factors \revised{are associated with} the assigned rank/position of the post on the r/popular feed?
    \item How is the post's rank on the r/popular feed \revised{associated with} the engagement on the post?
\end{enumerate}

RQ1 and RQ2 examine the factors that influence Reddit's ranking algorithm on r/popular. Specifically, what factors affect how long a post stays and where it is placed on the feed. RQ3 aims to quantify how algorithmic ranking decisions affect subsequent engagement on posts.

To answer these research questions, we capture a snapshot of the r/popular feed every 2 minutes over an 11-month period. Using over 1.5M consistently collected snapshots, we employ multiple regression analyses to examine the activity and movements (i.e., changes in position on the r/popular feed) of 10K posts from 694 distinct subreddits.


Through our analyses, we find that the total number of comments, along with recent commenting and voting activity, were the most predictive factors for a longer stay on the r/popular feed (RQ1) and for upward movement on the feed (RQ2). We also find that undesired comments \revised{(}i.e., toxic and moderator-removed comments\revised{)} were also predictive of a longer stay and upward movement, but to a lesser degree. Regarding how ranking on r/popular affects engagement (RQ3),  we find that posts which were higher on the feed received comments at a higher rate, as well as a greater proportion of undesired comments.

\revised{We employ multiple regression models to obtain relationships between post engagement and rank/position or movement on the r/popular feed. To complement these correlational insights, we employ a causal inference approach drawing on the potential outcomes framework~\cite{rubin2005causal} to assess causal relationships (detailed in Section~\ref{sec:causal-design}). This analysis provides evidence that increased commenting activity helps posts remain on r/popular longer (RQ1) and move up the feed (RQ2). We also find that upward movement on the feed leads to more commenting engagement (RQ3), including both overall and undesired engagement. However, we observe no disproportionate effect of rank on undesired engagement specifically.}

By systematically analyzing snapshots of r/popular, we provide insights into how a prominent, algorithmically curated trending feed made its ranking decisions and how those decisions may influence user engagement on the platform. Understanding how these opaque algorithms operate as an external party is inherently difficult. The internal workings of curation algorithms are proprietary and therefore inaccessible to users and researchers. This lack of access limits transparency regarding how user engagement is driven by content ranking and how the content users interact with is influenced by engagement. Our approach can empower users and researchers to examine the algorithmically curated feeds that determine what they interact with on social media.

\section{Related Work}

In this section, we review prior work on trending feeds/popularity, algorithmic curation systems, and algorithmic audits.

\subsection{Popularity \& Trending Feeds}

Popularity on social media has been studied before; however, these studies often focus on how users~\cite{devito-algorithmic-trap}, communities~\cite{increased-attention-github, my-paper}, or topics~\cite{twitter-trending-topics} change after becoming viral. For example, \citet{popularity-shocks-user} found that after users became viral, they increased their posting frequency and changed their posts to be similar to the post that made them viral. In general, the focus of these studies is the impact of popularity on (groups of) users. However, missing from prior work are investigations into why content or users became viral in the first place. \textit{To address this gap, we examine the factors that influence the systems that determine what is viral on social media, specifically the r/popular feed on Reddit.} It is important to understand how these algorithmic curation systems work because they are pervasive on social media and the popularity they induce can be a double-edged sword, particularly for those of marginalized identities~\cite{devito-algorithmic-trap}. Additionally, these systems can also propagate inflammatory content, as noted by mainstream media in recent times~\cite{inflammatory-content, radical-ideas}.

\subsection{Algorithmic Curation \& Ranking}

According to \citet{rader-gray-folk-theories}, algorithmic curation is the process of ``organizing, selecting, and presenting subsets of a corpus of information for consumptions.'' Within this definition, we are particularly focused on the organizing and presenting aspect because Reddit algorithmically ranks posts on r/popular. However, there exist other definitions~\cite{news-feed-fyi, congress-testimony, cura}. Algorithmic ranking is only one key mechanism in which algorithmic curation systems can exert power~\cite{diakopoulous-black-box}. By ranking content, specifically on social media feeds, these algorithms essentially have the power to determine what is important by prioritizing content or users over one another. To show how impactful algorithmic ranking can be, \citet{joachims-clicks} examined users' clickthrough behavior on Google's result page and found that participants' trust in Google's retrieval/ranking function led them to click on highly ranked links regardless of their quality or relevance to the query. Additionally, \citet{watts-music-markets} found that ranking songs in music markets by total downloads produced more unpredictability and inequality compared to groups who had songs ranked randomly. \textit{We extend this line of work by examining the influence of algorithmic ranking on social media trending feeds.}

\subsection{Algorithmic Audits}

Per \citet{auditing-algorithms-definition}, an \textit{algorithmic audit} is ``a method of repeatedly and systematically querying an algorithm with inputs and observing the corresponding outputs in order to draw inferences about its opaque inner workings.'' In this paper, we query the r/popular feed for 11 months and examine its ranking decisions to infer details about its internals and impact on engagement. Our study complements the large corpus of existing social media algorithm audits performed on platforms like X/Twitter~\cite{lower-higher-quality}, YouTube~\cite{radical-pathways-youtube, resnick-youtube-scrubbing}, Facebook~\cite{facebook-ideological-segregation}, and TikTok~\cite{tiktok-explanations} to name a few. Additionally, prior studies often focus on political bias that may be built into curation algorithms which differs from our focus on the factors that influence, and are influenced by, popularity.
\textit{We extend this line of work and conduct an audit of algorithmic ranking on a relatively underexplored site, Reddit's r/popular feed.}

To conduct these algorithmic audits, researchers commonly utilize sock-puppet accounts~\cite{twitter-sock-puppet-bias, diakopoulos-more-accounts, tiktok-explanations} that ``use code scripts to create simulated users''~\cite{resnick-youtube-scrubbing}. Sock-puppet accounts are necessary because platforms often do user-specific recommendations based on the user's activity. However, that raises the challenge of simulating realistic user behavior which is a limitation to these types of studies. \textit{By studying a trending feed that is available to everyone, we avoid having to use sock-puppet accounts.}
\section{Data} \label{sec:data}

To retrieve posts from the r/popular feed, we used PRAW---a Python wrapper for Reddit's API. Extending \citet{my-paper}'s methodology, every 2 minutes, we captured a \textit{snapshot} of the r/popular feed's top 100 posts from March 23, 2022, to February 8, 2023---approximately 11 months. 
A single API request returns at most 100 posts. Although the feed goes beyond the top 100, we settled on only requesting the top 100 posts to avoid having to make multiple requests for a single snapshot and to stay within Reddit API rate limits. Thus, during the study period, we made only one request every 2 minutes.

\subsection{Sampling Approach}

By taking snapshots of r/popular every 2 minutes, totaling 224,121 feed snapshots, we collected 134,661 unique posts from 1,423 distinct subreddits. \textit{For clarity, feed snapshots refer to a capture of the entire feed whereas a snapshot from now on refers to a capture of an individual post within a feed snapshot} of which there are 22,412,100 snapshots---100 per feed snapshot because there are 100 posts in a feed snapshot.

For tractability, we randomly sampled 10,000 posts from the 134,661 posts observed. The resulting sample contains 694 subreddits and 1,548,266 snapshots. For the rest of the paper, our findings are based on this representative sample of 10,000 posts and their respective snapshots. From the sample, we found that posts stay on r/popular for about 164 snapshots on average ($\mu=163.63$, $\sigma=140.25$). Additionally, posts from the sample, on average, stay on r/popular's top 100 for 6.1 hours ($\mu=6.11$, $\sigma=5.09$). Along with these snapshots, we \revised{use} Pushshift~\cite{pushshift} to obtain the comments for each post.

\subsection{Identifying Undesirable Activity} \label{sec:undesired-definition}

In recent years, news outlets have suggested that social media platforms are intentionally promoting antisocial content through algorithmic prioritization to drive greater user engagement~\cite{radical-ideas, inflammatory-content}. To investigate whether antisocial behavior has any interactions with algorithmic curation, specifically on r/popular, we \revised{employ} \citet{bert-model}'s fine-tuned BERT model, which assigns three toxicity-related scores for each comment: \textsc{non\_toxic}, \textsc{slightly\_toxic}, and \textsc{highly\_toxic}. Each score ranges from 0 to 1. If a comment \revised{exceeds} a score of 0.5---a threshold used in prior work~\cite{conversations-gone-alright}---for either \textsc{slightly\_toxic} or \textsc{highly\_toxic}, then we \revised{label} it as toxic. \citet{bert-model} fine-tuned the model using r/AskReddit comments and showed its generalizability to 99 other large subreddits. Since the subreddits that appear on r/popular are mostly large, we claim that using this model is well suited for identifying toxic comments made within our r/popular posts. Additionally, some comments were removed, presumably by a moderator or bot, before they could be archived by Pushshift. \textit{To better capture antisocial behavior, we combine comments that contained ``[removed]'' with the ones flagged by the BERT model under an umbrella term: ``undesired comments.''}

\subsection{Features} \label{sec:features}

In this section, we describe the features we \revised{use} for our regression models and provide a brief justification for inclusion. \textit{This feature list applies to posts captured at a specific snapshot/time.} Thus, the features are measured at a specific time, e.g., the number of comments at a particular snapshot.

The first feature is:

\begin{enumerate}
    \item \textit{Content Type:} Whether the post contains a link, video, image, or just text. This is a categorical variable where \textit{image posts} are the reference category, i.e., the category in which all other content types are compared to.
\end{enumerate}

Content type functions as a control variable to capture differences between posts that include images (49.19\%), links (16.19\%), text (12.45\%), and videos (22.17\%).

\begin{enumerate}
    \setcounter{enumi}{1}
    \item \textit{Rank:} Where the post is on r/popular where rank 1 is the top of the feed and rank 100 is the bottom.
    \item \textit{Age (hours):} The amount of time, in hours, since the post's creation.
\end{enumerate}

Rank is the focus of our audit and because r/popular emphasizes content that is \textit{currently} popular, the recency of a post is a natural feature to include.

The next set captures the activity within the post's thread, i.e., comment section. We also have features that utilized the labels produced by the BERT model described previously.

\begin{enumerate}
    \setcounter{enumi}{3}
    \item \textit{Num. Comments:} The total number of comments at the time the snapshot was taken.
    \item \textit{Recent Comments:} The number of comments made in the previous 10 minutes.
    \item \textit{Proportion Undesired:} The proportion of comments that were labeled as undesired comments.
    \item \textit{Proportion Recent Undesired:} The proportion of comments labeled as undesirable in the last 10 minutes.
    \item \textit{Score:} The number of upvotes on the post minus the number of downvotes.
    \item \textit{Recent Votes:} The number of votes the post received in the last 10 minutes.
    \item \textit{Proportion Upvotes:} The proportion of \textit{all} votes that were upvotes. 
\end{enumerate}

We \revised{include} recent activity because there is a ``hot'' calculation\textsuperscript{2} that likely includes comment and vote velocity, i.e., how many comments are coming in currently. \revised{We define recent activity as the activity in the past 10 minutes} because posts stay at a rank for an average 7.5 of minutes before moving onto another ($\mu=7.49$, $\sigma=7.77$). Calculating the proportion of undesired comments helps us test whether Reddit's ranking algorithm is in any way influenced by the presence of undesired activity.

\begin{enumerate}
    \setcounter{enumi}{10}
    \item \textit{Num. Subscribers:} The number of subscribers the post's origin subreddit has.
\end{enumerate}

Lastly, communities are integral to Reddit, which is why we \revised{include} their size as a control variable in the models.

Table~\ref{tab:descriptive} provides the geometric mean and standard deviations for each feature. We \revised{use} the geometric versions of these measures because the variables are log-transformed in our regression analyses. Additionally, the geometric standard deviations are used to inform the units we \revised{use} to scale the regression coefficients which are also shown in Table~\ref{tab:descriptive}. The consistency to use 2x for all units except the proportion of upvotes is to assist with interpretability.

\begin{table}[t]
    \centering
    \small
    \begin{tabularx}{\linewidth}{Xrrrr}
        \textbf{Feature} & \textbf{Unit} & $\mu$ & $\sigma$ & \textbf{Median} \\ \midrule
        Age (hours) & 2x & 8.738 & 1.644 & 9.067 \\
        Num. Comments & 2x & 745.569 & 3.130 & 864 \\
        Rec. Comments & 2x & 11.829 & 3.164 & 12 \\
        Prop. Undesired & 2x & 0.189 & 1.660 & 0.190 \\
        Rec. Prop. Undesired & 2x & 0.160 & 2.417 & 0.201 \\
        Score & 2x & 13.161K & 2.587 & 13.697K \\
        Rec. Votes & 2x & 357.792 & 3.021 & 396 \\
        Prop. Upvotes & 1.05x & 0.917 & 1.065 & 0.930 \\
        Num. Subscribers & 2x & 3.781M & 3.704 & 3.411M
    \end{tabularx}
    \caption{
        Descriptive statistics for the features used in the regression analyses where $\mu$ and $\sigma$ are the geometric mean and standard deviation, respectively. The regression results in the following sections are scaled using the ``Unit'' column informed by $\sigma$. `K' is for thousand,`M' is for million.
    }
    \label{tab:descriptive}
\end{table}

\section{\revised{Overarching Causal Inference Design}} \label{sec:causal-design}

\revised{To complement our regression analysis, we adopt a causal inference approach drawing on the potential outcomes framework~\cite{rubin2005causal}. Our goal is to examine the causal effects of a post receiving various ``treatments'' (e.g., moving up the feed) compared to what would have happened if the treatments were not administered (i.e., the ``counterfactuals''). Although our work relies on observational data, the potential outcomes framework allows us to estimate the missing ``counterfactual'' for each treated post based on the outcomes of similar (i.e., matched), observed posts that were not treated. In other words, matching enables a quasi-experimental setup to estimate counterfactuals while accounting for numerous covariates.}

\revised{We use this causal framework to conduct a time series analysis, matching on covariates measured in the snapshots leading up to the treatment. We employ stratified propensity score matching~\cite{psm-intro} to identify comparable control snapshots for each treatment. Following matching, we apply generalized linear models (GLMs) to estimate the effects of the treatments on outcomes measured in the snapshots immediately following the treatment.}

\subsection{\revised{Treatment Definitions}}

\revised{The treatments in our causal analysis include changes in the number of recent comments (\treatmentcomments{}), the proportion of undesired comments during that time (\treatmentundesired{}), and rank movements on the r/popular feed (\treatmentrank{}). We define treatment and control groups using consistent percentile-based thresholds, and we define weak and strong versions of each treatment.}

\revised{For recent comments, the weak treatment (\treatmentcommentsweak{}) is defined as receiving more comments in the past 10 minutes than the median ($> 11$ comments), and the strong treatment (\treatmentcommentsstrong{}) is defined as receiving at least the 75th percentile ($\geq 24$ comments). The control is defined as receiving at most the median ($\leq 11$ comments).}

\revised{For undesired comments, we define treatments based on the \textit{proportion} of undesired comments received in the past 10 minutes. We use the proportion of undesired comments to define our treatment instead of the number of undesired comments, as the latter would be confounded by the overall number of recent comments. The weak treatment (\treatmentundesiredweak{}) is defined as receiving a proportion of undesired comments greater than the median ($> 18.75\%$), the strong treatment (\treatmentundesiredstrong{}) is greater than or equal to the 75th percentile ($\geq 31.25\%$), and the control is less than or equal to the median ($\leq 18.75\%$).}

\revised{Finally, for rank movement, the weak treatment (\treatmentrankweak{}) is defined as moving at least one rank up the feed, and the strong treatment (\treatmentrankstrong{}) is defined as moving up by at least five ranks (75th percentile for rank movement). If the post does not climb the feed---i.e., either stays in the same position or falls---then it is considered a control sample.}

\revised{These cutoffs are summarized in Appendix~\ref{sec:appendix-matching}, which also includes information regarding filtering. For each treatment, we estimate its effect on outcomes of tenure (RQ1), rank movement (RQ2), and engagement (RQ3).}


\subsection{\revised{Stratified Propensity Score Matching}} \label{sec:matching}

\revised{For a quasi-experimental design, we conduct matching to find comparable treated and control groups. A propensity score model matches snapshots based on their likelihood of receiving the treatment, quantified as propensity scores. In particular, we conduct stratified propensity score matching---which balances the bias-variance tradeoff of comparing too biased (one-to-one matched) or too variant (unmatched) samples~\cite{smd-threshold-2}. This approach distributes groups with similar propensity scores into strata, and is a well-established method adopted in prior work~\cite{koustuv-ads, verma_examining_2022, yuan_mental_2023, charlotte-quasi}.}

\revised{For each treatment, we train an AdaBoost classifier to estimate propensity scores using the following covariates: (1) post age, (2) title text (top 200 bi-grams), (3) rank, (4) time at rank, (5) number of comments, (6) number of undesired comments, (7) overall comment rate, and (8) undesired comment rate. Snapshots are then sorted by their propensity scores and partitioned into equally sized strata, such that snapshots within a stratum are similarly likely to receive the treatment based on the covariates.}

\revised{To ensure high-quality strata, we exclude any stratum with fewer than 200 treatment or control snapshots, as well as strata where the standardized mean difference (SMD) between treatment and control groups exceeds 0.3---a common threshold in prior work~\cite{smd-threshold-1, smd-threshold-2}. This matching approach approximates the presence of a counterfactual snapshot that was not treated, enabling us to assess causal relationships between treatments and outcomes. Further details on the number of strata pre- and post-filtering are provided in Appendix~\ref{sec:appendix-matching}.}

\subsection{\revised{Generalized Linear Models (GLMs)}}

\revised{After filtering for high-quality strata, we employ GLMs and use Bayesian inference to estimate whether the treatment had any effect on the outcome of interest. We use a hierarchical Bayesian model to pool effects across strata, where within-stratum effects are sampled from a global distribution whose mean represents the average treatment effect.}

\revised{We present our findings with probability of direction ($pd$) and region of practical equivalence ($ROPE$) (full) as these are commonly used when communicating the strength and significance of findings in a Bayesian context~\cite{makowski2019indices}. The probability in $ROPE$ can be interpreted as the probability that a significant effect does \textit{not} exist, and $pd$ can be interpreted as the probability that an effect is in a particular direction. To facilitate discussion of whether the effect size is large enough to be ``significant,'' we define our $ROPE$ to be between -4.8\% and 5\%, and we consider effect sizes within the ROPE to be ``not significant.'' We also report the mean of the estimated effect size along with 95\% credible interval.}
\section{RQ1: Tenure on r/popular} \label{sec:rq1}

RQ1 asks what factors \revised{are associated with} how long a post stays on r/popular (i.e., tenure). To conduct this analysis, we built logistic regression models to estimate whether the post continues to exist on the feed in the next snapshot---approximately 2 minutes later. \revised{Additionally, we fit a set of GLMs to assess causal links between commenting activity and tenure.} This section describes the \revised{models} and \revised{their} findings in further detail.

\subsection{Logistic Regression}

Essentially, the task is to take an r/popular post during a snapshot and predict whether the same post continues to be on r/popular during the \textit{next snapshot}. Thus, factors that helped a post stay on the feed in the next snapshot also helped elongate its tenure on r/popular. Because we did not observe posts outside the top 100, we built three models for the top 50 (\mff{}), top 25 (\mtw{}), and top 10 (\mtn{}) to have enough observations of posts both inside and outside of these intervals. Additionally, testing multiple rank intervals helped examine the robustness of our results. Each respective model only includes posts that, at some point, reached its respective rank interval (see Table~\ref{tab:top-n} for the exact number of posts). For example, if a post $p$ is in $n$ snapshots before it reaches the top 50 and $m$ snapshots after, then \mff{} uses only the $m$ snapshots after its initial breakthrough.

\subsection{\revised{Causal Inference Analysis}}

\revised{We examined the causal relationship between commenting activity and post tenure by using a set of logistic regression models. We apply these models on our stratified data to estimate the effects of recent commenting activity (\treatmentcomments{}) and undesired activity (\treatmentundesired{}) on the odds of appearing within the top 50, 25, 10 ranks in the next snapshot.}


\subsection{Results}

\begin{table}[t]
    \centering
    \small
    \begin{tabular}{lrrr}
        \textbf{Feature} & \mff{} & \mtw{} & \mtn{}  \\ \midrule
        Link (vs. image)        & 88.93\%*          & 9.77\%*     & -18.32\%*                \\
        Text (vs. image)        & 47.98\%*          & 11.12\%*    & -2.21\%\enspace          \\
        Video (vs. image)       & -0.13\%\enspace   & -6.24\%*    & 12.01\%*                 \\ \midrule
        Age                     & -38.91\%*         & -39.16\%*   & -43.14\%*                \\
        Num. Comments           & 91.52\%*         & 105.37\%*   & 84.11\%*                \\
        Rec. Comments           & 19.08\%*          & 23.71\%*    & 27.76\%*                 \\
        Prop. Undesired         & 2.64\%*           & 5.26\%*     & 16.80\%*                 \\
        Prop. Rec. Undesired    & 3.18\%*           & 2.93\%*     & 3.94\%*                  \\
        Score                   & -47.89\%*         & -52.84\%*   & -57.45\%*                \\
        Rec. Votes              & 161.14\%*         & 125.10\%*   & 74.51\%*                 \\
        Prop. Upvotes           & -0.60\%\enspace   & 4.10\%*     & -4.72\%*                 \\
        Num. Subscribers        & 3.01\%*           & 2.39\%*     & 10.76\%*                 \\ \midrule
        Num. Posts              & 5,697             & 3,499       & 1,875                    \\ 
        Num. Snapshots          & 1,062,717         & 687,386     & 391,167                  \\ \midrule
        $R^2$                   & 0.230             & 0.261       & 0.285
    \end{tabular}
    \caption{
        Results from the three logistic regression models (\mff{}, \mtw{}, \mtn{}) predicting a post's tenure on the r/popular feed (*$p<0.05$, Bonferroni-adjusted). Percentages indicate the expected change in odds that a post will stay on the top 50 (\mff{}), 25 (\mtw{}), and 10 (\mtn{}) in the next snapshot---approximately 2 minutes later---given a unit increase in each respective feature (see Table \ref{tab:descriptive}). For example, a post with 2 times the number of comments as another post is expected to have 91.52\% greater odds of staying in the top 50.
    }
    \label{tab:top-n}
\end{table}

Table~\ref{tab:top-n} presents our findings from the three logistic regression models we built. The percentages in Table~\ref{tab:top-n} correspond to the change in odds of staying in the top 50, 25, or 10 when the respective feature increases multiplicatively by its standardized unit listed in Table~\ref{tab:descriptive}.

\textbf{Impact of Overall Engagement.} \revised{The} strongest factors \revised{associated with an} \revised{increase} in odds of staying near the top of the feed \revised{are} the number of total comments, recent comments, and recent votes. We also found age (i.e., time since post creation) to be a strong factor \revised{associated with} \revised{decreases in} the odds of staying near the top. These are consistent across all rank intervals and are natural for an engagement-based popular feed.

We also observed that increasing the score on a post \revised{corresponds} to a drastic decrease in the odds of staying on the feed. \revised{According to the $R^2$ values from each model, a substantial amount of variance remains unexplained. This suggests that unmodeled factors may be strongly correlated with score, contributing to the strong negative relationship observed in Table~\ref{tab:top-n}.}

\revised{From our causal analysis, we found that increasing the number of comments increased a post's tenure on r/popular in some cases. We found strong evidence that both the weak (\treatmentcommentsweak{}) and the strong (\treatmentcommentsstrong{}) treatments had a significant effect on the odds of appearing within the top 10 (0.23\% in ROPE), with a probability of $>$99.98\% ($pd$) that the effect was positive for the weak treatment (mean=17.59\%, CI[9.53\%, 25.86\%]), and a probability of 99.88\% ($pd$) that the effect was positive for the strong treatment (mean=28.15\%, CI[14.23\%, 45.06\%]). However, there was little evidence that either treatment significantly increased the odds of staying in the top 25 and top 50 ($\geq$18.45\% in ROPE). Thus, we were only able to causally link an increase in recent comments to tenure on the top 10.}

\textbf{Impact of Undesired Activity.} \revised{According to Table~\ref{tab:top-n}, we} found that increasing the proportion of undesired comments and the number of subscribers \revised{has} a comparatively small but significant \revised{positive relationship with} the odds that a post stays within the top 50 and top 25, with a stronger effect in the top 10. While these effects do not necessarily indicate that the ranking algorithm is intentionally designed to promote antisocial content to drive more user engagement, these results fail to rule it out. These \revised{relationships} also hint at an interaction between undesired comments and the rank of a post, and that the \revised{relationship with} undesired comments is strongest for posts at the top of the feed.

\revised{From our causal analysis, we found that increasing the proportion of undesired comments could increase a post's tenure on r/popular in some limited cases. We found evidence that the strong version of the treatment (\treatmentundesiredstrong{}) had a significant effect on the odds of appearing in the top 10 (2.58\% in ROPE), with a probability of 99.93\% ($pd$) that this effect is positive (mean=13.09\%, CI[5.34\%, 21.90\%]). However, we found little evidence that the weak treatment (\treatmentundesiredweak{}) had any significant effect on appearing within any rank threshold ($\geq$64.73\% in ROPE), or that the strong treatment had any significant effect on the odds of appearing the top 25 or top 50 ($\geq$88.70\% in ROPE).}

\section{RQ2: Rank on r/popular} \label{sec:rq2}

\begin{figure}[t]
    \centering
    \includegraphics[width=\linewidth]{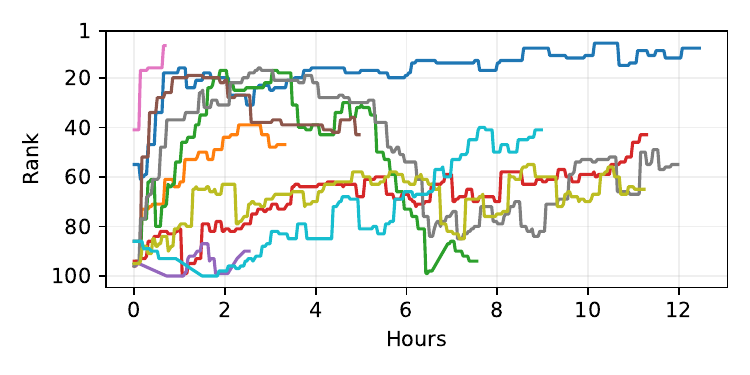}
    \caption{
        The rank trajectories of 10 r/popular posts where hour 0 is the first moment they reached the top 100. Note the stepwise movement as posts tended to jump to different ranks instead of gradually shifting between nearby ranks.
    }
    \label{fig:trajectories}
\end{figure}

In this section, we describe \revised{what factors are associated with a post's rank on r/popular}. 
As seen in Figure~\ref{fig:trajectories}, \revised{posts do} not gradually change between nearby ranks over time. Instead, the typical rank trajectory of a post consists of two parts: (1) where the post stays at one rank for several minutes, and (2) where the post suddenly jumps to another, distant rank. To characterize this discontinuous behavior, we used two separate regression models to determine which factors affect: (1) \textit{when} a post changes rank, and (2) \revised{\textit{how far} it moves} when it does.

\subsection{Multinomial Regression (Rank Movement)}

The first model is a multinomial regression that estimated whether a post will move up the feed, down the feed, or stay in its current position in the next snapshot. Posts tend to stay at the same rank for several minutes at a time (see Figure~\ref{fig:trajectories}), and this model helped us determine which factors accelerated the movement of a post, and in which direction.

\subsection{Ordinal Regression (Rank)}

\revised{We also aimed to estimate how far a post moves when it does, so we employed a second model:} an ordinal regression that estimated the post's next rank. Only snapshots where a post changed rank were considered so that we could get a more detailed picture of which rank a post changed to when it moved. We \revised{constructed a} latent-variable model, where a post's rank is determined by a continuous latent variable ($z$), and the observed rank depends on which cut points $z$ falls between. The model estimated both the cut points and the association between each feature and the expected value of $z$. The flexibility of the cut points within the model makes it well suited to handle the uneven ``distances'' between ranks---e.g., it might be less likely for a post to move from rank 2 to rank 1 than it is from rank 99 to 98. We implemented this model in a Bayesian framework with strongly informative priors on the cut points and weakly informative priors on the coefficients. We provided more details on the priors in Appendix~\ref{sec:model-2.2-appendix}.

\subsection{\revised{Causal Inference Analysis}}

\revised{We examined the causal relationship between commenting activity and rank movement using an approach similar to the earlier multinomial regression. We used a set of multinomial regression models on our stratified data to estimate the impacts of recent commenting activity (\treatmentcomments{}) and undesired activity (\treatmentundesired{}) on the odds of upward movement and downward movement against no movement in the next snapshot.}

\begin{table}[t]
    \centering
    \small
    \begin{tabularx}{\linewidth}{Xrr}
        \textbf{Feature} & \textbf{Up} & \textbf{Down} \\ \midrule
         Link (vs. image)           & 2.757\%\enspace       & 4.812\%*             \\
         Text (vs. image)           & 0.401\%\enspace       & 7.842\%*             \\
         Video (vs. image)          & 1.147\%\enspace       & -0.727\%\enspace     \\ \midrule
         Age                        & 4.117\%*              & 10.114\%*            \\
         Num. Comments              & -1.997\%*             & -7.200\%*            \\
         Rec. Comments              & 7.333\%*              & -4.127\%*            \\
         Prop. Undesired            & 0.751\%\enspace       & -2.681\%*            \\
         Prop. Rec. Undesired       & 0.285\%\enspace       & -0.801\%\enspace     \\
         Score                      & -11.626\%*            & 11.018\%*            \\
         Rec. Votes                 & 14.796\%*             & 3.886\%*             \\
         Prop. Upvotes              & 1.393\%*              & 0.803\%\enspace      \\
         Num. Subscribers           & 0.097\%\enspace       & -2.937\%*            \\ \midrule
         Num. Observations          & & 1,548,266                                  \\
         Nagelkerke $R^2$           & & 0.043
    \end{tabularx}
    \caption{
        Results from the multinomial regression model predicting when a post moves and in which direction (*$p<0.05$, Bonferroni-adjusted). Percentages represent expected change in odds of moving up or down the feed (compared to no movement) in the next snapshot, given a unit increase in each respective feature (see Table \ref{tab:descriptive}). For example, a post that is 2 times as old  as another post is expected to have 4.117\% greater odds of moving up the feed and 10.114\% lower odds of moving down the feed in next snapshot.
    }
    \label{tab:rank-movements}
\end{table}

\begin{figure}[t]
    \centering
    \includegraphics[width=\linewidth]{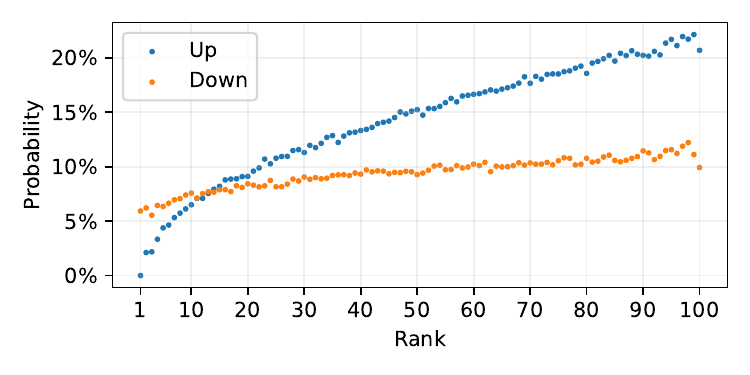}
    \caption{
        The probability of a post moving up or down on r/popular in the next snapshot across ranks estimated by a multinomial logistic regression. Note that most of the time a post did not move in the next snapshot.
    }
    \label{fig:rank-movements}
\end{figure}

\subsection{Results}

\begin{table}[t]
    \centering
    \small
    \begin{tabularx}{\linewidth}{Xr}
        \textbf{Feature} & \textbf{$\Delta z$} \\ \midrule
         Link (vs. image)           & 0.001\enspace\enspace     \\
         Text (vs. image)           & 0.033**                   \\
         Video (vs. image)          & -0.015*\enspace\          \\ \midrule
         Age                        & 0.163**                   \\
         Num. Comments              & -0.053**                  \\
         Rec. Comments              & -0.062**                  \\
         Prop. Undesired            & -0.021**                  \\
         Prop. Rec. Undesired       & -0.008**                  \\
         Score                      & 0.132**                   \\
         Rec. Votes                 & -0.081**                  \\
         Prop. Upvotes              & -0.018**                  \\
         Num. Subscribers           & -0.028**                  \\
    \end{tabularx}
    \caption{
        Results from the ordinal regression model estimating how far a post jumps in rank. Values represent expected change in the latent variable $z$ representing rank on r/popular (Figure \ref{fig:cut-points} shows the cut points which map $z$ to ranks) given a unit increase to a feature (see Table \ref{tab:descriptive}). For example, a post that is 2 times older than another post is expected to have a 0.163 increase in $z$ in the next snapshot where the rank changes. Parameters whose \textit{probability of direction}, a Bayesian analog of $p$-values~\cite{makowski2019indices}, are greater than 0.95 are denoted with a single asterisk (*), and those greater than 0.999 are denoted with two (**).
    }
    \label{tab:next-rank}
\end{table}

\begin{figure}
    \centering
    \includegraphics[width=\linewidth]{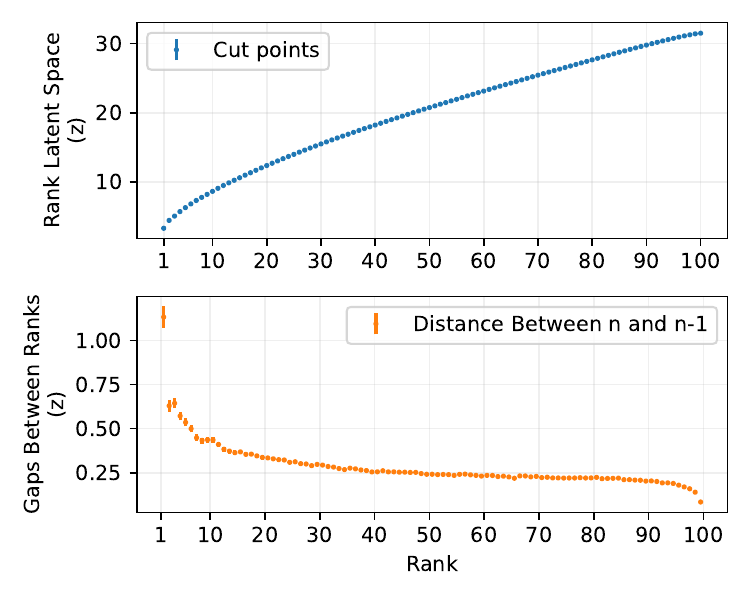}
    \caption{
        The ordinal regression model's estimated cut points for each rank and the distances between adjacent ranks in the latent space used to represent rank. The independent feature associations in Table~\ref{tab:rank-movements} correspond to this latent space.
    }
    \label{fig:cut-points}
\end{figure}

Starting with the multinomial regression, Table~\ref{tab:rank-movements} shows \revised{the associations between} a \revised{twofold} increase to a feature\revised{---}excluding the proportion of upvotes\revised{---}and the probability of a post moving up or down the feed. Additionally, Figure~\ref{fig:rank-movements} shows the baseline probabilities for each rank movement across \revised{ranks}. Note that the probability of moving up and down do not sum to 100\% \revised{as the remainder reflects the likelihood that the post remains in the same position in the next snapshot.}

\textbf{\revised{Association Between} Rank \revised{\&} Movement.} We found that posts closer to the top of the feed moved less frequently and less far. Figure~\ref{fig:rank-movements} shows that the probability that a post moves in either direction reduces as its rank gets closer to 1, and Figure~\ref{fig:cut-points} that shows the gaps between cut points close to rank 1 are wider than cut points closer to rank 100. This is consistent with the intuition that there would be less competition toward the top of the feed as it would be rare for posts to achieve that level of ``quality.''

\textbf{Impact of Overall Activity on Rank.} Similar to our results in RQ1, the number of recent comments and recent votes \revised{are} strong factors \revised{associated with increases in} the probability \revised{moving} up in rank, as seen in Table~\ref{tab:rank-movements}. In terms of how far a post \revised{moves}, Table~\ref{tab:next-rank} indicates that the total number of comments, recent comments, and recent votes \revised{have} a similar magnitude \revised{relationship}. However, Table~\ref{tab:rank-movements} \revised{shows} that the total number of comments tend\revised{s} to ``stabilize'' the rank of a post, decreasing the odds of movement in either direction, but still favoring upward movement. Similarly, the number of subscribers \revised{is not associated with an} increase \revised{in} probability of upward movement directly, but it still favored upward movement by decreasing the probability of downward movement (-2.937\%). These findings are consistent with an engagement-based ranking algorithm, but they highlight the influence of recent activity.

\revised{Our causal analysis found that increasing the number of comments caused posts to move up the feed by simultaneously increasing the odds of moving up the feed and decreasing the odds of moving down the feed. Specifically, we found strong evidence that the receiving more than 11 recent comments (\treatmentcommentsweak{}) had a significant effect on the odds of moving upward ($<$0.03\% in $ROPE$), and a $>$99.98\% probability ($pd$) that this effect is positive (mean=10.63\%, CI[8.65\%, 12.52\%]). We also found strong evidence that receiving more than 10 recent comments (\treatmentcommentsweak{}) had a significant effect on the odds of moving downward ($<$0.03\% in $ROPE$), and a $>$99.98\% probability ($pd$) that this effect is negative (mean=-11.57\%, CI[-13.67\%, -9.79\%]).}

\revised{We found similar results with the strong treatment (\treatmentcommentsstrong{}), though with slightly lower confidence due to the relative lack of suitable strata (see Appendix~\ref{sec:appendix-matching}). Specifically, we found evidence that receiving at least 24 recent comments (\treatmentcommentsstrong{}) had a significant effect ($<$4.53\% in $ROPE$) on the odds of moving upward, and a 99.58\% probability ($pd$) that this effect is positive (mean=11.40\%, CI[3.15\%, 19.48\%]). We also found strong evidence that the same treatment (\treatmentcommentsstrong{}) had a significant effect on the odds of moving downward (0.1\% in ROPE), and a $>$99.98\% probability ($pd$) that this effect is negative (mean=-17.80\%, CI[-24.19\%, -10.60\%]).}


\textbf{Impact of Undesired Activity on Rank.} Table~\ref{tab:rank-movements} \revised{shows} that the proportion of undesired comments slightly \revised{associated with a decrease in} the likelihood of moving down the feed, and Table~\ref{tab:next-rank} indicates that undesired comments \revised{has} a relatively small \revised{but positive relationship with how far a post moves up the feed (negative in terms of $z$)}. \revised{This is in contrast with Table~\ref{tab:rank-movements} that only identifies a negative relationship between undesired comments and posts moving downwards.}

\revised{However, from our causal analysis, we found little evidence that the proportion of undesired comments has any significant effect on rank movement up or down, with a probability in $ROPE$ of 97.98\% or greater for both the weak (\treatmentundesiredweak{}) and strong versions of the treatment (\treatmentundesiredstrong{})}.
\section{RQ3: Engagment on r/popular} \label{sec:rq3}

Our analyses from RQ1 and RQ2 provided empirical insights into the \revised{factors are associated with rank} on r/popular. Finally, for RQ3, we investigated how \textit{rank} on the r/popular feed \revised{influences} the engagement on posts, specifically the frequency of comments gained before the next snapshot.

\subsection{Negative Binomial Regression}

\revised{To estimate the relationship between each feature in Section~\ref{sec:features} and the comment rate on a post during the next snapshot---measured as the number of \textit{new} comments between two snapshots---we trained a single negative binomial regression to estimate both the rate of non-undesired comments and undesired comments by using an interaction term that indicates whether the model is estimating the former or the latter. Thus, by default, the model measures the association between each feature and the rate of non-undesired comments and the interaction term measures whether there is any difference in the relationship when estimating the rate of undesired comments instead.}

\subsection{\revised{Causal Inference Analysis}}

\revised{To estimate whether moving up the feed \textit{causally} increases the rate of comments overall, undesired comments, or the proportion of undesired comments, we trained a set of negative binomial regression models. These models utilize our stratified data to estimate the impact moving up the feed one rank (\treatmentrankweak{}) and five ranks (\treatmentrankstrong{})---i.e., the 75th percentile---on commenting activity. More specifically, we measure each treatment's effect on the overall commenting rate, undesired commenting rate, and proportion of undesired comments during the period before the next snapshot.}

\subsection{Results}

\begin{table}[t]
    \centering
    \small
    \begin{tabularx}{\linewidth}{Xrrr}
        \textbf{Feature} & \textbf{$\neg$Und.} & \multicolumn{1}{r}{\textbf{Und:$\neg$Und.}} & \multicolumn{1}{r}{\textbf{Net Und.}} \\         \midrule
        Link (vs. image)        & -3.440\%*          & +0.342\%             & -3.110\%     \\
        Self (vs. image)        & -2.179\%*          & +0.650\%             & -1.543\%     \\
        Video (vs. image)       & 0.924\%*           & +0.006\%             & 0.930\%      \\ \midrule
        Age                     & -13.140\%*         & -0.467\%\enspace     & -13.545\%    \\
        Num. Comments           & 22.331\%*          & -2.601\%*            & 19.149\%     \\
        Rec. Comments           & 61.740\%*          & +1.056\%*            & 63.449\%     \\
        Prop. Undesired         & -14.467\%*         & +115.633\%*          & 84.437\%     \\
        Prop. Rec. Und.         & -1.500\%\enspace   & +8.713\%*            & 7.082\%      \\
        Score                   & -13.992\%*         & +2.062\%*            & -12.218\%    \\
        Rec. Votes              & 13.530\%*          & -3.422\%*            & 9.645\%      \\
        Prop. Upvotes           & 3.519\%*           & -1.077\%*            &  2.405\%     \\
        Num. Subscribers        & 1.414\%*           & -0.344\%*            & 1.066\%      \\ \midrule
        Pseudo $R^2$            & 0.29 \enspace\enspace &                   &
    \end{tabularx}
    \caption{
        Results from the negative binomial regression model predicting the rate of non-undesired and undesired comments (*$p<0.05$, Bonferroni-adjusted). Percentages in ``$\neg$Und.'' denote expected change in non-undesired comments given a unit increase in the respective feature (see Table \ref{tab:descriptive}). Percentages in ``Und:$\neg$Und.'' denote expected change in the ratio of undesired to non-undesired comments gained. Net expected change to undesired comments is shown in the ``Net Und.'' column. For example, a post that is 2 times older than another is expected to have 13.1\% fewer non-undesired comments and a further 0.47\% fewer undesired comments, for a net of 13.5\% fewer undesired comments.
    }
    \label{tab:engagement}
\end{table}

\begin{figure}[t]
    \centering
    \includegraphics[width=\linewidth]{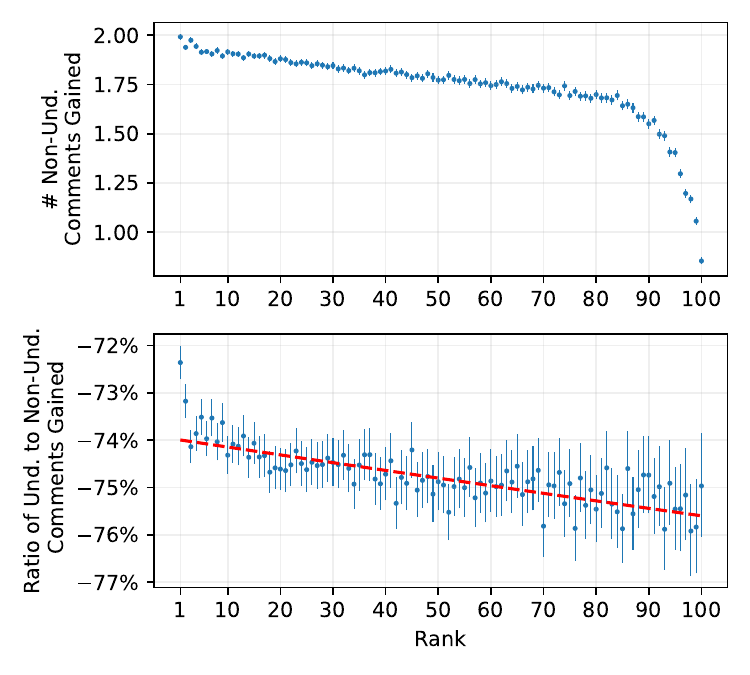}
    \caption{
        Plots of the intercepts for non-undesired comments (top) and undesired comments (bottom). Error bars indicate 95\% confidence intervals. Assuming a ``standard'' post with features equal to $\mu$ (see Table~\ref{tab:descriptive}), the top plot can be interpreted as the expected number of non-undesired comments gained in 2 minutes at a given rank, and the bottom plot can be interpreted as the expected ratio of undesired comments to non-undesired comments gained at a given rank. For example, a ``standard'' post at rank 1 is expected to gain about 2 non-undesired comments every 2 minutes, and is expected to gain 72.4\% fewer undesired comments than non-undesired comments. The bottom plot also shows a decreasing trend in the ratio of undesired comments with increasing rank, at a rate of about 0.064\% per rank (red dashed line).
    }
    \label{fig:engagement-rank-coefficients}
\end{figure}

Table~\ref{tab:engagement} shows how our feature set from Section~\ref{sec:features}\revised{---}excluding rank which is visualized in Figure~\ref{fig:engagement-rank-coefficients}\revised{---}is associated with the rate of non-undesired comments and the rate of undesired comments.

\textbf{\revised{Association Between} Engagement \revised{\&} Non-Undesired Activity.} From Table~\ref{tab:engagement}, we found that the number of comments, recent commenting activity, and recent voting activity \revised{have} the strongest associations with future non-undesired activity. Additionally, the \revised{relationship} \textit{recent} commenting activity \revised{has with} non-undesired commenting rate (61.74\%) \revised{is} greater than the one found with the comments \revised{overall} (22.33\%). This hints that the active discussions \revised{plays} a more important role to future engagement than the total number of discussions that have already occurred.

\textbf{\revised{Association Between} Engagement \revised{\&} Undesired Activity.} For undesired activity, we found that the proportion of undesired comments \revised{has} the strongest association (84.43\%) out of all the ones found in Table~\ref{tab:engagement}. This \revised{potentially} suggests that having a greater proportion of undesired activity \revised{invites} similarly undesired activity in the future.

\textbf{Impact of Rank on Commenting Activity.} Regarding rank's associations with commenting rate, we found that the rate of non-undesired comments \revised{falls} fairly consistently from ranks 2 through 80, visualized in the first subplot in Figure~\ref{fig:engagement-rank-coefficients}. Afterward, the rate of non-undesired comments \revised{falls} precipitously below rank 80.

\revised{From our causal analysis, we found strong evidence that moving up the feed increases the comment rate experienced by the post. Both our weak (\treatmentrankweak{}) and strong (\treatmentrankstrong{}) treatments increase commenting rate. Specifically, the weak treatment (i.e., moving up the feed at least one rank) had significant (0.03\% in $ROPE$) mean effect of 8.00\% (CI[5.76\%, 10.41\%]) with a $>$99.98\% probability ($pd$) that this effect is positive. Similarly, for our strong treatment (i.e., moving up the feed at least five ranks) had a significant (0.05\% in $ROPE$) mean effect of 10.19\% (CI[7.14\%, 13.09\%]) with a $>$99.98\% probability ($pd$) that this effect is positive. Thus, we found strong evidence that higher positions on the feed incur more comments.}

\textbf{Impact of Rank on Undesired Activity.} From the second subplot in Figure~\ref{fig:engagement-rank-coefficients}, we observed that the ratio of undesired comments \revised{falls} as the post is placed lower on the feed. Specifically, the ratio of undesired comments to non-undesired comments \revised{falls} at a rate of 0.064\% per rank ($p < 0.05$). Additionally, ranks 2 and 3 may indicate that there is a sharp rise in the ratio at the peak of the r/popular feed, however, it is difficult to be sure given the confidence intervals.


\revised{Although Figure~\ref{fig:engagement-rank-coefficients} shows a slight increase in the proportion of undesired comments at higher ranks, our causal analysis found no \textit{disproportionate} increase in undesired comments relative to overall commenting. Both treatments had a statistically significant (0.1\% in $ROPE$) effect on the number of undesired comments: 8.65\% mean for \treatmentrankweak{} (CI[6.29\%, 11.18\%]) and 10.74\% mean for \treatmentrankstrong{} (CI[7.57\%, 13.77\%]). However, these effect sizes are comparable to those observed for overall commenting. When analyzing the proportion of undesired comments---thus controlling for the overall comment volume---we found no significant effect ($>$99.98\% in $ROPE$). This suggests that the increase in undesired comments likely coincides with a broader increase in overall engagement, rather than indicating a targeted or disproportionate effect.}

\section{Discussion}

Here, we discuss the implications of our methodology and findings for future audits of algorithmically curated feeds. 

\subsection{Empowerment Through Transparency}

Prior literature has shown that people are often unaware that algorithms control what they see and, in turn, how their content is spread~\cite{rader-explanations-facebook, silas-awareness}. This unawareness can lead to harm~\cite{feedvis-was-not-close} resulting in calls for greater transparency regarding these algorithmic systems~\cite{alvarado-algorithmic-experience}. Our study enables transparency by quantifying how factors like commenting rate and undesirable activity affect algorithmic ranking on a prominent social media feed. 

\textbf{Implications for Content Creators.} Our study has implications for content creators, particularly those whose livelihoods are tied to algorithmic decision making. The relationship we identified between algorithmic rank and engagement highlights how the algorithms employed by social media platforms essentially get to decide which content is prioritized and which is not. This has led content creators on platforms like YouTube to blindly attempt to appease algorithms by adjusting their content creation process~\cite{creator-friendly-algorithms}. Audits of algorithmic feeds can empower creators by providing accurate information about how their actions can affect content prioritization on their respective platforms.

\textbf{Implications for Content Consumption.} We found that the order in which content is ranked can influence the levels and types of user engagement within algorithmically curated feeds. Similarly, the types of content that are recommended and where they are positioned can affect users' content consumption practices~\cite{reranking, serps-apis}. Prior work has examined this impact through users' trust in Google's ranking algorithm, which leads to a greater number of clicks on top-ranked links regardless of their accuracy or relevance to the query~\cite{joachims-clicks}. This \textit{trust bias} that \citet{joachims-clicks} refer to, coupled with the increased engagement levels observed in our analysis (see Section~\ref{sec:rq3}), can catalyze the spread of misinformation---especially if higher ranks within feeds are misconstrued as an indicator of content quality and trustworthiness. To alleviate some of these pitfalls, some platforms have provided explanations for their algorithmic recommendations, however, \citet{tiktok-explanations} found that those on TikTok in particular were generic and inapplicable to the user. Further research is needed to empirically 
understand how algorithmic rank affects users' perceptions of content highlighted on social media feeds.

\subsection{Implications for Content Moderation}

Prior work has shown that sudden popularity can be disruptive to moderation teams that have to manage increased activity, particularly from newcomers~\cite{eternal-september, chan2022community, my-paper}.
Despite how integral algorithmic curation is to social media platforms, moderators currently have little influence on algorithmic curation systems. This lack of agency impedes their ability to direct their communities and highlight desirable content---an idea that has been explored in prior work~\cite{fred-creator-hearts}. Given that moderators have little influence on these systems, there is a need for design interventions that provide moderators with more controls or ways to ``override'' algorithmic curation systems. These additional affordances can be used to highlight desirable content in feeds, which has theoretical foundations~\cite{successful-communities-book} and is currently done with existing mechanisms such as awards, upvotes, pins, and flairs~\cite{charlotte-positive-reinforcement}. These additional controls may take the form of various sliders that adjust latent weights on the feed to emphasize different priorities (e.g., increase user diversity on the feed), visual indicators for moderator-assigned contributions, or specific feeds that filter for moderator-selected contributions---similar to the ones the New York Times has to filter for reader- and editor-selected comments~\cite{nyt-comment-picks}.

\subsection{Implications for Future Empirical Audits} \label{sec:future-audits}

Despite the pervasiveness of algorithmic curation on social media, studying it as external researchers is exceedingly difficult. This difficulty is compounded by data access restrictions as platforms lock down their APIs, e.g., X/Twitter~\cite{twitter-api-access} and Reddit~\cite{reddit-admin-pushshift}. The reduction in API accessibility has led researchers to use other methods like sock-puppet accounts~\cite{post-api-era, twitter-sock-puppet-bias, diakopoulos-more-accounts, resnick-youtube-scrubbing, tiktok-explanations} \revised{and} data donations~\cite{jhaver-data-donation, chouaki-data-donation} that each have their own set of drawbacks. These restrictions have led to the reliance on externally-maintained datasets like the now defunct Pushshift---which has been heavily used in prior research~\cite{shagun-explanations, atcheson-new-members, renkai-pushshift, tal-august-summaries}. \textit{To avoid these challenges, we developed a robust pipeline to collect large-scale high-fidelity snapshots of Reddit's trending feed r/popular which can be adapted for other platforms with similar feeds.}

However, we recognize that these approaches are extremely time- and resource-intensive, posing significant challenges for scalability without adequate financial backing and collaborative support. Although partnerships between academia and industry have yielded important research outcomes, such collaborations are increasingly rare. Given these circumstances, and echoing a theme identified in a 2024 CCC Workshop Visioning Report~\cite{ccc-workshop}, this presents an opportunity for researchers to share auditing infrastructures and data-sharing protocols that support the collection, storage, and access of social media data for research purposes. For example, in the industry circuit, \textit{data clean rooms}~\cite{herbrich2022data} are emerging for several companies to have a shared and secure data infrastructure. However, the creation of these infrastructures must also adhere to relevant privacy and legal standards such as GDPR~\cite{gdpr} or CCPA~\cite{ccpa}, especially for sensitive data. This would enable more research like ours, informing future regulation, while safeguarding the rights and privacy of social media users.
\subsection{Limitations \revised{\& Future Directions}} \label{sec:limitations}

We recognize that our study bears limitations, however, these limitations also suggest interesting future directions.


\revised{Our findings are shaped by the demographics of Reddit users. Applying them elsewhere requires careful attention to demographic similarities between Reddit and the platform in question. Our study employed a robust causal inference design that mitigated potential confounds and led to stronger inferences than purely correlational methods. However, we cannot cannot claim ``true causality,'' due to the lack of ``true counterfactuals'' and the potential for unobserved confounders. Future research can build on our work through controlled experimental studies to further validate and extend these insights.}

The relatively low $R^2$ our models exhibited indicate that there are other factors outside of the ones we have included that influence the outcomes. One such factor may include details about surrounding posts on the r/popular feed. A hint that our model is incomplete is the observation that increasing the score on a post corresponds to a drastic decrease in the odds of staying on the feed, as found in RQ1 and RQ2. This counterintuitive result may be due to a correlation between a post's score and some unobserved influence. Regardless, future work can build upon these models by including more factors about surrounding posts that are competing for the same finite number of spots on these feeds. Furthermore, future work could also employ more advanced statistical methods like a stochastic transitivity model~\cite{johnson_bayesian_2013}.

Lastly, because our findings were based on observational data, future work will have to examine more user-centered effects of algorithmic ranking---similar to \citet{watts-music-markets, joachims-clicks}, but on social media.
Specifically, future work could design controlled experiments to examine how algorithmic rank influences individual users and shapes their perceptions of highly ranked content. These studies can also incorporate interviews or surveys to investigate user perceptions of content quality and relevance, as well as affective responses (e.g., trust), and how these factors relate to user engagement and online content consumption.
Together, these studies can offer complementary insights that provide a more comprehensive understanding of algorithmic ranking on social media platforms.
\section{Conclusion}

Now that algorithmic curation has become integral to online ecosystems, examining it and its effects on user behavior becomes critical. In this paper, we conducted a comprehensive empirical audit of one such system: Reddit's r/popular feed. Through this, we successfully quantified how recent commenting rate and other factors influence algorithmic ranking on r/popular, as well as how rank/position on r/popular \revised{impact} subsequent engagement and undesirable behavior. Our findings are based on millions of snapshots of the top 100 ranks on r/popular consistently collected over 11 months, an approach that can be applied to other platforms. Additionally, we discussed the implications of our findings for stakeholders, including content creators, highlighted moderators' lack of agency in community curation systems, and proposed future research directions on algorithmic curation amid reduced data access. All in all, studying algorithmic curation goes back to user agency: should users have control over how they interact with content and how their content is distributed online? To address this, we must first understand how these systems are embedded in our online environments, thus enabling us to make informed decisions about governing them.

\fontsize{9pt}{8pt}{\selectfont\bibliography{bib}}

\begin{thebibliography}{65}
\providecommand{\natexlab}[1]{#1}

\bibitem[{Almerekhi, Kwak, and Jansen(2022)}]{bert-model}
Almerekhi, H.; Kwak, H.; and Jansen, B.~J. 2022.
\newblock Investigating toxicity changes of cross-community redditors from 2 billion posts and comments.
\newblock \emph{PeerJ Computer Science}.

\bibitem[{Alvarado and Waern(2018)}]{alvarado-algorithmic-experience}
Alvarado, O.; and Waern, A. 2018.
\newblock Towards Algorithmic Experience: Initial Efforts for Social Media Contexts.
\newblock In \emph{CHI}.

\bibitem[{Atcheson, Koshy, and Karahalios(2024)}]{atcheson-new-members}
Atcheson, A.; Koshy, V.; and Karahalios, K. 2024.
\newblock Not What it Used to Be: Characterizing Content and User-base Changes in Newly Created Online Communities.
\newblock In \emph{CHI}.

\bibitem[{August et~al.(2024)August, Lo, Smith, and Reinecke}]{tal-august-summaries}
August, T.; Lo, K.; Smith, N.~A.; and Reinecke, K. 2024.
\newblock Know Your Audience: The benefits and pitfalls of generating plain language summaries beyond the ``general'' audience.
\newblock In \emph{CHI}.

\bibitem[{Austin(2011)}]{psm-intro}
Austin, P.~C. 2011.
\newblock An {Introduction} to {Propensity} {Score} {Methods} for {Reducing} the {Effects} of {Confounding} in {Observational} {Studies}.
\newblock \emph{Multivariate Behavioral Research}.

\bibitem[{Bandy and Diakopoulos(2021)}]{diakopoulos-more-accounts}
Bandy, J.; and Diakopoulos, N. 2021.
\newblock More Accounts, Fewer Links: How Algorithmic Curation Impacts Media Exposure in Twitter Timelines.
\newblock \emph{PACM HCI}.

\bibitem[{Bao et~al.(2021)Bao, Wu, Zhang, Chandrasekharan, and Jurgens}]{conversations-gone-alright}
Bao, J.; Wu, J.; Zhang, Y.; Chandrasekharan, E.; and Jurgens, D. 2021.
\newblock Conversations Gone Alright: Quantifying and Predicting Prosocial Outcomes in Online Conversations.
\newblock In \emph{WWW}.

\bibitem[{Bartley et~al.(2021)Bartley, Abeliuk, Ferrara, and Lerman}]{twitter-sock-puppet-bias}
Bartley, N.; Abeliuk, A.; Ferrara, E.; and Lerman, K. 2021.
\newblock Auditing Algorithmic Bias on Twitter.
\newblock In \emph{WebSci}.

\bibitem[{Baumgartner et~al.(2020)Baumgartner, Zannettou, Keegan, Squire, and Blackburn}]{pushshift}
Baumgartner, J.; Zannettou, S.; Keegan, B.; Squire, M.; and Blackburn, J. 2020.
\newblock The Pushshift Reddit Dataset.
\newblock arXiv:2001.08435.

\bibitem[{Calma(2023)}]{twitter-api-access}
Calma, J. 2023.
\newblock Scientists say they can’t rely on {Twitter} anymore.
\newblock \emph{The Verge}.

\bibitem[{CCPA(2018)}]{ccpa}
CCPA. 2018.
\newblock California {Consumer} {Privacy} {Act} ({CCPA}).
\newblock \emph{State of California - Department of Justice - Office of the Attorney General}.

\bibitem[{Chan et~al.(2022)Chan, Atreyasa, Chancellor, and Chandrasekharan}]{chan2022community}
Chan, J.; Atreyasa, A.; Chancellor, S.; and Chandrasekharan, E. 2022.
\newblock Community Resilience: Quantifying the Disruptive Effects of Sudden Spikes in Activity within Online Communities.
\newblock {CHI EA}.

\bibitem[{Chan et~al.(2024)Chan, Lambert, Choi, Chancellor, and Chandrasekharan}]{my-paper}
Chan, J.; Lambert, C.; Choi, F.; Chancellor, S.; and Chandrasekharan, E. 2024.
\newblock Understanding Community Resilience: Quantifying the Effects of Sudden Popularity via Algorithmic Curation.
\newblock In \emph{ICWSM}.

\bibitem[{Choi et~al.(2025)Choi, Lambert, Koshy, Pratipati, Do, and Chandrasekharan}]{fred-creator-hearts}
Choi, F.; Lambert, C.; Koshy, V.; Pratipati, S.; Do, T.; and Chandrasekharan, E. 2025.
\newblock Creator Hearts: Investigating the Impact Positive Signals from YouTube Creators in Shaping Comment Section Behavior.
\newblock In \emph{CHI}.

\bibitem[{Choi et~al.(2023)Choi, Kang, Lee, and Kim}]{creator-friendly-algorithms}
Choi, Y.; Kang, E.~J.; Lee, M.~K.; and Kim, J. 2023.
\newblock Creator-friendly Algorithms: Behaviors, Challenges, and Design Opportunities in Algorithmic Platforms.
\newblock In \emph{CHI}.

\bibitem[{Chouaki et~al.(2024)Chouaki, Chakraborty, Goga, and Zannettou}]{chouaki-data-donation}
Chouaki, S.; Chakraborty, A.; Goga, O.; and Zannettou, S. 2024.
\newblock What News Do People Get on Social Media? Analyzing Exposure and Consumption of News through Data Donations.
\newblock In \emph{WWW}.

\bibitem[{Cotter, Cho, and Rader(2017)}]{news-feed-fyi}
Cotter, K.; Cho, J.; and Rader, E. 2017.
\newblock Explaining the News Feed Algorithm: An Analysis of the ``News Feed FYI''' Blog.
\newblock In \emph{CHI EA}.

\bibitem[{DeVito(2022)}]{devito-algorithmic-trap}
DeVito, M.~A. 2022.
\newblock How Transfeminine TikTok Creators Navigate the Algorithmic Trap of Visibility Via Folk Theorization.
\newblock \emph{PACM HCI}.

\bibitem[{Diakopoulos(2013)}]{diakopoulous-black-box}
Diakopoulos, N. 2013.
\newblock Algorithmic accountability reporting: On the investigation of black boxes.
\newblock \emph{Tow Center for Digital Journalism}.

\bibitem[{Eckles(2022)}]{congress-testimony}
Eckles, D. 2022.
\newblock Algorithmic transparency and assessing effects of algorithmic ranking.

\bibitem[{Epstein and Robertson(2015)}]{seme-politics-candidates}
Epstein, R.; and Robertson, R.~E. 2015.
\newblock The search engine manipulation effect (SEME) and its possible impact on the outcomes of elections.
\newblock \emph{PNAS}.

\bibitem[{Eslami et~al.(2024)Eslami, Gilbert, Schoenebeck, Baumer, Chandrasekharan, Mooy, Karahalios, Karger, Cottom, Monroy-Hernández, Terveen, and Wihbey}]{ccc-workshop}
Eslami, M.; Gilbert, E.; Schoenebeck, S.; Baumer, E. P.~S.; Chandrasekharan, E.; Mooy, M.~D.; Karahalios, K.; Karger, D.; Cottom, T.~M.; Monroy-Hernández, A.; Terveen, L.; and Wihbey, J. 2024.
\newblock The Future of Research on Social Technologies: CCC Workshop Visioning Report.
\newblock arXiv:2404.10897.

\bibitem[{Eslami et~al.(2015)Eslami, Rickman, Vaccaro, Aleyasen, Vuong, Karahalios, Hamilton, and Sandvig}]{feedvis-was-not-close}
Eslami, M.; Rickman, A.; Vaccaro, K.; Aleyasen, A.; Vuong, A.; Karahalios, K.; Hamilton, K.; and Sandvig, C. 2015.
\newblock ``I always assumed that I wasn't really that close to [her]'': Reasoning about Invisible Algorithms in News Feeds.
\newblock In \emph{CHI}.

\bibitem[{GDRP(2023)}]{gdpr}
GDRP. 2023.
\newblock Data {Sharing}: {A} {Code} of {Practice}.
\newblock \emph{Information Commissioner's Office}.

\bibitem[{Gillespie(2016)}]{trending-is-trending}
Gillespie, T. 2016.
\newblock \#trendingistrending: when algorithms become culture.
\newblock In \emph{Algorithmic {Cultures}}. Routledge.

\bibitem[{González-Bailón et~al.(2023)González-Bailón, Lazer, Barberá, Zhang, Allcott, Brown, Crespo-Tenorio, Freelon, Gentzkow, Guess, Iyengar, Kim, Malhotra, Moehler, Nyhan, Pan, Rivera, Settle, Thorson, Tromble, Wilkins, Wojcieszak, de~Jonge, Franco, Mason, Stroud, and Tucker}]{facebook-ideological-segregation}
González-Bailón, S.; Lazer, D.; Barberá, P.; Zhang, M.; Allcott, H.; Brown, T.; Crespo-Tenorio, A.; Freelon, D.; Gentzkow, M.; Guess, A.~M.; Iyengar, S.; Kim, Y.~M.; Malhotra, N.; Moehler, D.; Nyhan, B.; Pan, J.; Rivera, C.~V.; Settle, J.; Thorson, E.; Tromble, R.; Wilkins, A.; Wojcieszak, M.; de~Jonge, C.~K.; Franco, A.; Mason, W.; Stroud, N.~J.; and Tucker, J.~A. 2023.
\newblock Asymmetric ideological segregation in exposure to political news on {Facebook}.
\newblock \emph{Science}.

\bibitem[{Gurjar et~al.(2022)Gurjar, Bansal, Jangra, Lamba, and Kumaraguru}]{popularity-shocks-user}
Gurjar, O.; Bansal, T.; Jangra, H.; Lamba, H.; and Kumaraguru, P. 2022.
\newblock Effect of popularity shocks on user behaviour.
\newblock In \emph{ICWSM}.

\bibitem[{He et~al.(2023)He, Gordon, Popowski, and Bernstein}]{cura}
He, W.; Gordon, M.~L.; Popowski, L.; and Bernstein, M.~S. 2023.
\newblock Cura: Curation at Social Media Scale.
\newblock \emph{PACM HCI}.

\bibitem[{Herbrich(2022)}]{herbrich2022data}
Herbrich, T. 2022.
\newblock Data Clean Rooms.
\newblock \emph{Computer Law Review International}.

\bibitem[{Hsu et~al.(2020)Hsu, Vaccaro, Yue, Rickman, and Karahalios}]{silas-awareness}
Hsu, S.; Vaccaro, K.; Yue, Y.; Rickman, A.; and Karahalios, K. 2020.
\newblock Awareness, Navigation, and Use of Feed Control Settings Online.
\newblock In \emph{CHI}.

\bibitem[{Jhaver, Bruckman, and Gilbert(2019)}]{shagun-explanations}
Jhaver, S.; Bruckman, A.; and Gilbert, E. 2019.
\newblock Does Transparency in Moderation Really Matter? User Behavior After Content Removal Explanations on Reddit.
\newblock \emph{PACM HCI}.

\bibitem[{Jhaver et~al.(2023)Jhaver, Garimella, De~Choudhury, Wilson, Vashistha, and Mitra}]{jhaver-data-donation}
Jhaver, S.; Garimella, K.; De~Choudhury, M.; Wilson, C.; Vashistha, A.; and Mitra, T. 2023.
\newblock Getting Data for CSCW Research.
\newblock In \emph{CSCW Companion}.

\bibitem[{Joachims et~al.(2017)Joachims, Granka, Pan, Hembrooke, and Gay}]{joachims-clicks}
Joachims, T.; Granka, L.; Pan, B.; Hembrooke, H.; and Gay, G. 2017.
\newblock Accurately Interpreting Clickthrough Data as Implicit Feedback.
\newblock \emph{SIGIR}.

\bibitem[{Johnson and Kuhn(2013)}]{johnson_bayesian_2013}
Johnson, T.~R.; and Kuhn, K.~M. 2013.
\newblock Bayesian {Thurstonian} models for ranking data using {JAGS}.
\newblock \emph{Behavior Research Methods}.

\bibitem[{Kampenes et~al.(2007)Kampenes, Dybå, Hannay, and Sjøberg}]{smd-threshold-1}
Kampenes, V.~B.; Dybå, T.; Hannay, J.~E.; and Sjøberg, D. I.~K. 2007.
\newblock A systematic review of effect size in software engineering experiments.
\newblock \emph{Information and Software Technology}.

\bibitem[{Kiciman, Counts, and Gasser(2018)}]{smd-threshold-2}
Kiciman, E.; Counts, S.; and Gasser, M. 2018.
\newblock Using longitudinal social media analysis to understand the effects of early college alcohol use.
\newblock In \emph{ICWSM}.

\bibitem[{Kiene, Monroy-Hern\'{a}ndez, and Hill(2016)}]{eternal-september}
Kiene, C.; Monroy-Hern\'{a}ndez, A.; and Hill, B.~M. 2016.
\newblock Surviving an "Eternal September": How an Online Community Managed a Surge of Newcomers.
\newblock In \emph{CHI}.

\bibitem[{Kou et~al.(2024)Kou, Ma, Zhang, Zhou, and Gui}]{renkai-pushshift}
Kou, Y.; Ma, R.; Zhang, Z.; Zhou, Y.; and Gui, X. 2024.
\newblock Community Begins Where Moderation Ends: Peer Support and Its Implications for Community-Based Rehabilitation.
\newblock In \emph{CHI}.

\bibitem[{Kraut and Resnick(2012)}]{successful-communities-book}
Kraut, R.~E.; and Resnick, P. 2012.
\newblock \emph{Building successful online communities: Evidence-based social design}.
\newblock MIT Press.

\bibitem[{Kulshrestha et~al.(2017)Kulshrestha, Eslami, Messias, Zafar, Ghosh, Gummadi, and Karahalios}]{political-search-bias}
Kulshrestha, J.; Eslami, M.; Messias, J.; Zafar, M.~B.; Ghosh, S.; Gummadi, K.~P.; and Karahalios, K. 2017.
\newblock Quantifying Search Bias: Investigating Sources of Bias for Political Searches in Social Media.
\newblock In \emph{CSCW}.

\bibitem[{Lambert, Choi, and Chandrasekharan(2024)}]{charlotte-positive-reinforcement}
Lambert, C.; Choi, F.; and Chandrasekharan, E. 2024.
\newblock ``Positive reinforcement helps breed positive behavior''': Moderator Perspectives on Encouraging Desirable Behavior.
\newblock \emph{PACM HCI}.

\bibitem[{Lambert, Saha, and Chandrasekharan(2025)}]{charlotte-quasi}
Lambert, C.; Saha, K.; and Chandrasekharan, E. 2025.
\newblock Does Positive Reinforcement Work?: A Quasi-Experimental Study of the Effects of Positive Feedback on Reddit.
\newblock In \emph{CHI}.

\bibitem[{Liu, Wu, and Resnick(2024)}]{resnick-youtube-scrubbing}
Liu, A.; Wu, S.; and Resnick, P. 2024.
\newblock How to Train Your YouTube Recommender to Avoid Unwanted Videos.
\newblock In \emph{ICWSM}.

\bibitem[{Makowski et~al.(2019)Makowski, Ben-Shachar, Chen, and L{\"u}decke}]{makowski2019indices}
Makowski, D.; Ben-Shachar, M.~S.; Chen, S.~A.; and L{\"u}decke, D. 2019.
\newblock Indices of effect existence and significance in the Bayesian framework.
\newblock \emph{Frontiers in psychology}.

\bibitem[{Maldeniya et~al.(2020)Maldeniya, Budak, Robert~Jr., and Romero}]{increased-attention-github}
Maldeniya, D.; Budak, C.; Robert~Jr., L.~P.; and Romero, D.~M. 2020.
\newblock Herding a Deluge of Good Samaritans: How GitHub Projects Respond to Increased Attention.
\newblock In \emph{WWW}.

\bibitem[{Metaxa et~al.(2021)Metaxa, Park, Robertson, Karahalios, Wilson, Hancock, Sandvig et~al.}]{auditing-algorithms-definition}
Metaxa, D.; Park, J.~S.; Robertson, R.~E.; Karahalios, K.; Wilson, C.; Hancock, J.; Sandvig, C.; et~al. 2021.
\newblock Auditing algorithms: Understanding algorithmic systems from the outside in.
\newblock \emph{Foundations and Trends{\textregistered} in Human--Computer Interaction}.

\bibitem[{Mousavi, Gummadi, and Zannettou(2024)}]{tiktok-explanations}
Mousavi, S.; Gummadi, K.~P.; and Zannettou, S. 2024.
\newblock Auditing algorithmic explanations of social media feeds: A case study of tiktok video explanations.
\newblock In \emph{ICWSM}.

\bibitem[{O'Sullivan(2019)}]{inflammatory-content}
O'Sullivan, D. 2019.
\newblock How fake accounts pushing inflammatory content went viral – with the help of {YouTube}’s algorithms {\textbar} {CNN} {Business}.
\newblock \emph{CNN}.

\bibitem[{Perriam, Birkbak, and Freeman(2020)}]{post-api-era}
Perriam, J.; Birkbak, A.; and Freeman, A. 2020.
\newblock Digital methods in a post-{API} environment.
\newblock \emph{International Journal of Social Research Methodology}.

\bibitem[{Piccardi et~al.(2024)Piccardi, Saveski, Jia, Hancock, Tsai, and Bernstein}]{reranking}
Piccardi, T.; Saveski, M.; Jia, C.; Hancock, J.; Tsai, J.~L.; and Bernstein, M.~S. 2024.
\newblock Reranking Social Media Feeds: A Practical Guide for Field Experiments.
\newblock arXiv:2406.19571.

\bibitem[{Pogacar et~al.(2017)Pogacar, Ghenai, Smucker, and Clarke}]{seme-medicine}
Pogacar, F.~A.; Ghenai, A.; Smucker, M.~D.; and Clarke, C.~L. 2017.
\newblock The Positive and Negative Influence of Search Results on People's Decisions about the Efficacy of Medical Treatments.
\newblock In \emph{ICTIR}.

\bibitem[{Poudel and Weninger(2024)}]{serps-apis}
Poudel, A.; and Weninger, T. 2024.
\newblock Navigating the Post-API Dilemma.
\newblock In \emph{WWW}.

\bibitem[{Rader, Cotter, and Cho(2018)}]{rader-explanations-facebook}
Rader, E.; Cotter, K.; and Cho, J. 2018.
\newblock Explanations as Mechanisms for Supporting Algorithmic Transparency.
\newblock In \emph{CHI}.

\bibitem[{Rader and Gray(2015)}]{rader-gray-folk-theories}
Rader, E.; and Gray, R. 2015.
\newblock Understanding User Beliefs About Algorithmic Curation in the Facebook News Feed.
\newblock In \emph{CHI}.

\bibitem[{Reddit(2023)}]{reddit-admin-pushshift}
Reddit. 2023.
\newblock Reddit {Data} {API} {Update}: {Changes} to {Pushshift} {Access}.

\bibitem[{Ribeiro et~al.(2020)Ribeiro, Ottoni, West, Almeida, and Meira}]{radical-pathways-youtube}
Ribeiro, M.~H.; Ottoni, R.; West, R.; Almeida, V. A.~F.; and Meira, W. 2020.
\newblock Auditing radicalization pathways on YouTube.
\newblock In \emph{FAT*}.

\bibitem[{Rubin(2005)}]{rubin2005causal}
Rubin, D.~B. 2005.
\newblock Causal inference using potential outcomes: Design, modeling, decisions.
\newblock \emph{Journal of the American Statistical Association}.

\bibitem[{Saha et~al.(2021)Saha, Liu, Vincent, Chowdhury, Neves, Shah, and Bos}]{koustuv-ads}
Saha, K.; Liu, Y.; Vincent, N.; Chowdhury, F.~A.; Neves, L.; Shah, N.; and Bos, M.~W. 2021.
\newblock Advertiming matters: Examining user ad consumption for effective ad allocations on social media.
\newblock In \emph{CHI}.

\bibitem[{Salganik, Dodds, and Watts(2006)}]{watts-music-markets}
Salganik, M.~J.; Dodds, P.~S.; and Watts, D.~J. 2006.
\newblock Experimental study of inequality and unpredictability in an artificial cultural market.
\newblock \emph{Science}.

\bibitem[{Schlessinger et~al.(2023)Schlessinger, Garimella, Jakesch, and Eckles}]{twitter-trending-topics}
Schlessinger, J.; Garimella, K.; Jakesch, M.; and Eckles, D. 2023.
\newblock Effects of Algorithmic Trend Promotion: Evidence from Coordinated Campaigns in Twitter’s Trending Topics.
\newblock In \emph{ICWSM}.

\bibitem[{Verma et~al.(2022)Verma, Bhardwaj, Aledavood, De~Choudhury, and Kumar}]{verma_examining_2022}
Verma, G.; Bhardwaj, A.; Aledavood, T.; De~Choudhury, M.; and Kumar, S. 2022.
\newblock Examining the impact of sharing {COVID}-19 misinformation online on mental health.
\newblock \emph{Scientific Reports}.

\bibitem[{Wang et~al.(2024)Wang, Huang, Zhou, and Metaxa}]{lower-higher-quality}
Wang, S.; Huang, S.; Zhou, A.; and Metaxa, D. 2024.
\newblock Lower Quantity, Higher Quality: Auditing News Content and User Perceptions on Twitter/X Algorithmic versus Chronological Timelines.
\newblock \emph{PACM HCI}.

\bibitem[{Wang and Diakopoulos(2022)}]{nyt-comment-picks}
Wang, Y.; and Diakopoulos, N. 2022.
\newblock Highlighting High-quality Content as a Moderation Strategy: The Role of New York Times Picks in Comment Quality and Engagement.
\newblock \emph{TSC}.

\bibitem[{Wu(2019)}]{radical-ideas}
Wu, K.~J. 2019.
\newblock Radical ideas spread through social media. {Are} the algorithms to blame?
\newblock \emph{PBS}.

\bibitem[{Yuan et~al.(2023)Yuan, Saha, Keller, Isomets\"{a}, and Aledavood}]{yuan_mental_2023}
Yuan, Y.; Saha, K.; Keller, B.; Isomets\"{a}, E.~T.; and Aledavood, T. 2023.
\newblock Mental Health Coping Stories on Social Media: A Causal-Inference Study of Papageno Effect.
\newblock In \emph{WWW}.

\end{thebibliography}

\section{Ethics Statement} \label{sec:ethics}

We acknowledge that curation algorithms are often kept secret not only to safeguard intellectual property but also to prevent manipulation by users seeking undue attention or prioritization. Although our findings improve our understanding of these systems, we do not believe that they provide actionable strategies for malicious actors to compromise trending feeds, such as r/popular. Instead, by examining these systems, our work promotes transparency by assessing the factors that influence these systems. This analysis is valuable to a variety of stakeholders, as outlined earlier.

Regarding consent, the data used in this study was collected via a publicly accessible API and did not require Institutional Review Board (IRB) approval from our institution(s). To protect user privacy, no identities have been disclosed and we are not releasing the dataset. However, this study is fully replicable by collecting data snapshots using Reddit's API (see Section~\ref{sec:data}). The dataset is securely stored on a firewalled, password-protected server at our institution(s). Additionally, the content of the comments used were not revealed and were only used to label them as undesired or non-undesired using an offline instance of \citet{bert-model}'s BERT model.

\appendix

\section{Bayesian Priors for Model 2.2} \label{sec:model-2.2-appendix}

\begin{figure}[ht]
    \centering
    \includegraphics[width=\linewidth]{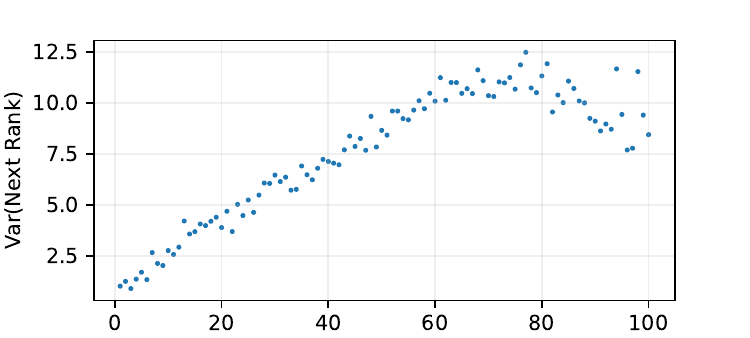}
    \caption{
Variance of a post's rank in the next snapshot (y-axis) plotted against a post's rank in the current snapshot (x-axis).}
    \label{fig:next-rank-variance}
\end{figure}

Our intuition is that it is ``easier'' to move between ranks further down the feed (e.g., from 95 to 91) than it is to move between ranks higher up the feed (e.g., from 5 to 1).
Indeed, based on Figure~\ref{fig:next-rank-variance}, we saw that the variance of rank in the next snapshot appears to increase approximately linearly with the rank in the current snapshot, up until rank 60-80. Beyond that, the variance appears to decrease due to the fact that ranks beyond 100 were censored from our dataset.
We used this observation to set strongly informative priors on our cut points. 

First, we assumed that the variance of next rank increases linearly with the current rank, then the standard deviation would increase proportionally to the square root of the current rank. In other words, this means that posts tend to traverse more ranks the further down the feed, and the number of ranks they typically traverse increases proportionally to the square root of the current rank.
We can capture this notion in the priors for our cut points by making the gap between cut points inversely proportional to the square root of rank. Thus we chose the following priors for our cut points $\kappa_k$ for each rank $k$:

$$
    \kappa_{k} - \kappa_{k-1} \sim \text{LogNormal}(\mu=\frac{\alpha}{\sqrt{k}},\sigma=0.1)
$$

$\alpha$ is a scaling factor with a weakly informative prior of $\alpha \sim \text{Exponential}(\lambda = 1)$, and $\kappa_1$ is fixed to $\kappa_1 = \alpha$.  

We also assume that a post's next rank will be close to its current rank. Thus, the intercepts for each current rank $k$ were given weakly informative priors of $\beta_{\text{rank}=k} \sim \text{Normal}(\mu=\kappa_{k}, \sigma=1)$. 

Finally, the remaining coefficients were given weakly informative priors: $\beta \sim \text{Normal}(\mu=0, \sigma=1).$

\onecolumn

\section{Stratified Propensity Score Matching} \label{sec:appendix-matching}

\begin{table*}[h]
    \centering
    \small
    \begin{tabular}{llllrrrr}
        \textbf{} & \textbf{} & \textbf{} & \textbf{} & 
        \multicolumn{2}{c}{\textbf{Strata}} & 
        \multicolumn{2}{c}{\textbf{Samples}} \\
        \cmidrule(r){5-6} \cmidrule(l){7-8}
        \textbf{RQ} & \textbf{Variable} & \textbf{Strength} & \textbf{Cutoff (Control, Treatment)} & 
        \textbf{Pre-Filter} & \textbf{Post} & \textbf{Per Stratum} & \textbf{Total} \\ \midrule
        1 \& 2 & Rec. Comments & Weak & $\leq Q(0.5)$ (11), $>Q(0.5)$ & 200 & 97 & 7,333 & 711,297 \\
        1 \& 2 & Rec. Comments & Strong & $\leq Q(0.5)$ (11), $\geq Q(0.75)$ (24) & 200 & 14 & 5,679 & 79,506 \\
        1 \& 2 & Rec. Und. & Weak & 0, $\geq$1 & 200 & 124 & 7,348 & 911,174 \\
        1 \& 2 & Rec. Und. & Strong & 0, $\geq Q(0.75)$ (5) & 200 & 39 & 2,283 & 153,342 \\
        1 \& 2 & Prop. Und. & Weak & $\leq Q(0.5)$ (18.75\%), $>Q(0.5)$ & 200 & 166 & 6,989 & 1,160,062 \\
        1 \& 2 & Prop. Und. & Strong & $\leq Q(0.5)$ (18.75\%), $\geq Q(0.75)$ (31.25\%) & 200 & 89 & 5,260 & 468,167 \\
        3 & Rank Movement & Weak & $\geq$0 (down), $<$0 (up) & 100 & 93 & 2,521 & 234,510 \\
        3 & Rank Movement & Strong & $\geq$0 (down), $\leq Q(0.25)$ (up 3) & 100 & 87 & 1,672 & 147,283 \\
    \end{tabular}
    \caption{Summary of the stratified propensity score matching approach adopted to estimate causal effects on the r/popular feed. $Q(n)$ denotes the percentile cutoff used to binarize treatment and control snapshots---e.g., $Q(0.5)$ is the 50th percentile. The number following $Q(n)$ denotes the absolute number or percentage resulting from that percentile cutoff---e.g., in the first row it takes 11 recent comments to reach the 50th percentile.}
    \label{tab:stratified-matching}
\end{table*}

\twocolumn
\section{GLMs/Hierarchical Bayesian Models} \label{sec:causal-models}

Here, we detail the models used for our causal inference analyses in Sections 5-7.

\subsection{RQ1}

Recall that we tested multiple treatments and outcomes, with four treatments (\treatmentcommentsweak, \treatmentcommentsstrong, \treatmentundesiredweak, \treatmentundesiredstrong), and three outcomes (top 10, 20, and 50). Each treatment/outcome pair was modeled separately for a total of $4 \times 3=12$ separate models. 

For a particular treatment, let $x_i=1$ if the $i$-th observation was in the treatment condition, and $x_i=0$ if it was in the control condition. Then, for a particular outcome $Y$, the probability $P(Y_{i}|x_{i})=p_{i}$ that a post would appear in the top 10 (20, or 50) in the next snapshot was modeled as follows:
$$ \text{logit}( p_{i} ) = \alpha_{s[i]} + \beta_{s[i]}\ x_{i}$$
Here, $s[i]$ denotes the stratum that the $i$-th observation was assigned to, and with separate intercepts $\alpha_{j}$ and coefficients $\beta_{j}$ for each stratum. Each intercept $\alpha_{j}$ represents the baseline (i.e., control) log-odds of appearing in the top 10 (20, or 50) within the $j$-th stratum. Each coefficient $\beta_{j}$ corresponds to the treatment effect in the $j$-th stratum. To compute the average treatment effect across all strata, the coefficients $\beta_j$ were sampled from a hyper-distribution with mean $\mu_{\beta}$ and standard deviation $\sigma_{\beta}$ (the intercepts were pooled in a similar manner):
$$ \beta_{j} \sim \text{Normal}(\mu_{\beta}, \sigma_{\beta})$$
The scalar $\mu_{\beta}$ thus represents the average treatment effect. Weakly informative priors were placed on these hyper-parameters (with similar priors on the hyper-parameters for the intercepts):
$$\mu_{\beta} \sim \text{Normal}(0, 1)$$
$$\sigma_{\beta} \sim \text{HalfCauchy}(\beta=1)$$
A posterior distribution for the average treatment effect $\mu_\beta$ was sampled for each treatment/outcome pair with Monte-Carlo Markov Chain sampling using PyMC. \footnote{https://github.com/pymc-devs/pymc}

\subsection{RQ2}

Again, recall that we tested multiple conditions and outcomes, with the same four treatment as before, and three outcomes (movement up, movement down, no movement). Each treatment was modeled separately, but the outcomes were modeled jointly using multinomial logistic regressions, yielding a total of just 4 models, one for each treatment.  

Let the possible values of the outcome $Y$ be up, none, and down. For each observation $i$, let $\frac{P(Y_{i}=\text{up}|x_{i})}{P(Y_{i}=\text{none}|x_{i})} = o_{\text{up}, i}$ be the odds that the post would move up the feed in the next snapshot against the not moving in the next snapshot, and similarly for $o_{\text{down}, i}$. We modeled these odds as follows:
$$ \log(o_{\text{up},i}) = \alpha_{\text{up},s[i]} + \beta_{\text{up}, s[i]}\  x_{i}$$
$$ \log(o_{\text{down},i}) = \alpha_{\text{down},s[i]} + \beta_{\text{down}, s[i]}\  x_{i}$$
Again, $s[i]$ denotes the stratum that the $i$-th observation was assigned to, and with separate intercepts ($\alpha_{\text{up},j}$ and  $\alpha_{\text{down},j}$) and coefficients ($\beta_{\text{up}, j}$, $\beta_{\text{down}, j}$) for each stratum. Each intercept $\alpha_{\text{up}, j}$ represents the baseline (i.e., control) log-odds of moving up the feed against no movement within the $j$-th stratum. Each coefficient $\beta_{\text{up}, j}$ corresponds to the treatment effect in the $j$-th stratum. The similar applies to $\alpha_{\text{down}, j}$  and $\beta_{\text{down}, j}$. The average treatment effect was modeled in a similar way as above, with similar priors on the hyper-parameters. Finally, posterior distributions for average treatment effects $\mu_{\beta_{\text{up}}}$ and $\mu_{\beta_{\text{down}}}$ were sampled for each treatment.

\subsection{RQ3}

For RQ3, we combined consecutive snapshots where there was no movement in rank into a single observation, yielding observations that spanned varying amounts of time, and which had rank movements just prior and just after the spanned time window. For the first two parts, where we investigated the impact on the number of overall comments and the number of undesired comments, we followed a similar procedure: Let $\Delta t_{i}$ be the time spanned during the $i$-th observation window, and let $Y_i$ be the number of comments received during that window. We modeled $Y_i$ using a negative binomial distribution with dispersion parameter $\alpha$:
$$ Y_{i} \sim \text{NBinom}(\mu=\lambda_{i}\ \Delta t_{i}, \alpha) $$
$$ \log(\lambda_{i}) = \gamma_{s[i]} + x_{i}\ \beta_{s[i]} $$
Here, $\lambda_{i}$ can be interpreted as the predicted commenting rate for the $i$-th observation, which is multiplied by $\Delta t_{i}$ to get the expected number of comments over that time window. Again, $s[i]$ denotes the stratum that the $i$-th observation was assigned to, and with separate intercepts $\gamma_{j}$ and coefficients $\beta_{j}$ for each stratum. Each intercept $\gamma_{j}$ represents the baseline (i.e. control) log-odds of commenting rate within the $j$-th stratum. Each coefficient $\beta_{j}$ corresponds to the treatment effect in the $j$-th stratum. The average treatment effect was modeled in a similar way as above, with similar priors on the hyper-parameters. Finally, posterior distributions for average treatment effect $\mu_{\beta}$ were sampled for each treatment.

For the third part, where we investigated the impact on the proportion of undesired comments, we followed a slightly different procedure. Let $Y_i$ be the number of undesired comments and let $n_{i}$ be the total number of comments received in the $i$-th observation window. We modeled $Y_i$ as follows:
$$ Y_{i} \sim \text{NBinom}(\mu=p_{i}\ n_{i}, \alpha) $$$$ \text{logit}(p_{i}) = \gamma_{s[i]} + x_{i}\ \beta_{s[i]} $$
Here, $p_i$ can be interpreted as the predicted proportion of undesired comments out of all comments received in the $i$-th observation. Each intercept $\gamma_{j}$ represents the baseline log ratio of undesired comments, and each coefficient $\beta_j$ corresponds to the treatment effect. Average treatment effects $\mu_{\beta}$ for each treatment were computed in a similar manner as before.

\end{document}